\documentstyle[12pt]{article}
\begin{document}
\def\eq#1{(\ref{#1})}
\newcommand{\rf}[1]{(\ref{#1})}
\def\VEV#1{\langle {#1}\rangle}
\newcommand{\beq}{\begin{equation}}
\newcommand{\eeq}{\end{equation}}
\newcommand{\bea}{\begin{eqnarray}}
\newcommand{\eea}{\end{eqnarray}}
\newcommand{\nn}{\nonumber \\}
\newcommand{\om}{\omega}
\newcommand{\el}{\Gamma^{\rm el}_W}
\newcommand{\tot}{\Gamma_{W}^{\rm tot}}
\newcommand{\ann}{\Gamma_{ann}}
\newcommand{\la}{\langle}
\newcommand{\ra}{\rangle}
\newcommand{\cm}{{\rm cm}}
\renewcommand{\min}{{\rm min}}
\newcommand{\G}{{\rm G}}
\newcommand{\yr}{{\rm yr}}
\newcommand{\gal}{L_{\rm Gal}}
\newcommand{\seed}{B_{\rm seed}}
\newcommand{\qcd}{T_{\rm QCD}}
\newcommand{\lw}{{L_W}}
\newcommand{\hh}{\tilde{H}_{\perp}}

\newcommand{\ew}{T_{\rm EW}}
\newcommand{\ns}{B_{\rm NS}}
\newcommand{\GeV}{{\rm GeV}}
\newcommand{\MeV}{{\rm MeV}}
\newcommand{\eV}{{\rm eV}}
\renewcommand{\H}{\bar{H}}
\def\gsim{\raise.3ex\hbox{$>$\kern-.75em\lower1ex\hbox{$\sim$}}}
\def\lsim{\raise.3ex\hbox{$<$\kern-.75em\lower1ex\hbox{$\sim$}}}
\newenvironment{marray}{\begin{equation}\begin{array}}%
{\end{array}\end{equation}}
%
%
\newenvironment{carray}{\begin{equation}\begin{array}{rcl}}%
{\end{array}\end{equation}}
\def\be{\@ifnextchar[{\def\ee{\end{equation}}\begin{equation}\l@b}%
{\def\ee{$$}$$}}
\def\l@b[#1]{\label{#1}}
\def\ba{\@ifnextchar[{\def\ee{\end{carray}}\begin{carray}\l@b}%
{\def\ee{\end{array}$$}$$\begin{array}{rcl}}}

\thispagestyle{empty}
\begin{titlepage}
\today

{\ \hfill HU-TFT-94-26}\\
\vskip 0.3cm\begin{center}
\LARGE
{\bf Dirac neutrinos and primordial magnetic fields}
\end{center}
\normalsize
\vskip1cm
\begin{center}
{\bf Kari Enqvist}\footnote{\em enqvist@phcu.helsinki.fi; on a leave of
absence from
NORDITA, Blegdamsvej 17, DK-2100 Copenhagen,
Denmark}
\\ \narrower{
Research Institute for Theoretical Physics, University of Helsinki,
P.O. Box 9, SF-00014 Helsinki, Finland}\\
and\\
{\bf A.I. Rez\footnote{\em rez@charley.izmiran.rssi.ru;
 rez@izmiran.troitsk.su} and V.B. Semikoz\footnote{\em
semikoz@charley.izmiran.rssi.ru; semikoz@izmiran.troitsk.su}}
\end{center}
\vskip-10pt
\narrower{
The
Institute of the Terrestrial Magnetism, the Ionosphere and Radio Wave
Propagation of the Russian Academy of Sciences, IZMIRAN,Troitsk, Moscow
region, 142092, Russia.}

\begin{abstract}\noindent
We consider random primordial magnetic fields and discuss their
dissipation, coherence length $L_0$,
scaling behaviour and constraints implied by the primoridal
nucleosynthesis.
Such magnetic fields could excite the right-helicity states
of Dirac neutrinos, with adverse consequences for nucleosynthesis.
We present solutions to the spin kinetic equation
of a Dirac neutrino traversing a random magnetic field in the cases
of large and small $L_0$, taking also into account elastic collisions.
Depending on the scaling behaviour and on the magnetic coherence length,
the lower limit on the neutrino magnetic moment
thus obtained could be as severe as
$10^{-20}\mu_B$.

\end{abstract}
\vfill
\end{titlepage}

\newpage
\section{Introduction}
The phase transitions of the very early universe may have generated
primordial magnetic fields, which could play an important role
in cosmology.  Large primordial fields can survive to this day
and provide a seed field for the galactic dynamo mechanism \cite{dynamo}
which amplifies the seed to produce the observed galactic
magnetic fields.
Several suggestions have been made as to the
possible mechanisms which could produce large primordial fields \cite{lpf}.
For instance, large fluctuations at the electroweak
phase transition might be responsible \cite{Vachaspati}, and it
has been argued \cite{eo} that after a proper statistical averaging one
could actually obtain seed fields of the required magnitude \cite{brand},
about $10^{-18}$ G. A more exotic possibility which also seems to
work is based on the observation that in Yang-Mills theories the
vacuum may have a permanent magnetic field, which is imprinted
on the comoving plasma already at the GUT scale \cite{poul}.

Primordial nucleosynthesis is sensitive to
magnetic fields, which modify both the Hubble rate and the
rates of the reactions that help
to build up the  light elements
\cite{Cheng}. Primordial nucleosynthesis may also be
affected in another way \cite{EOS} provided neutrinos are Dirac
particles. In that case
the right-helicity component of the neutrino
can be excited and brought into thermal equilibrium by scattering
of the left-helicity neutrinos off
the magnetic field, thus changing the effective number of degrees
of freedom.

In a recent estimate of the primordial helium abundance
\cite{OS} the allowed number of the additional
neutrino species was reduced to $\Delta N_{\nu} = N_{\nu} - 3 \simeq
0.1$. There is some uncertainty in this estimate due to the unkown magnitude
of systematic errors in the observed abundances, but even allowing
for very conservative systematic errors,
$\Delta N_\nu$ is definitely less than 1. Thus the coupling of a Dirac
neutrino to a primordial magnetic field should be weak enough
not to equilibrate the right-helicity states below the QCD
phase transition temperatures. This gives rise to a bound on
the combination $\mu_\nu B$, where $\mu_\nu$ is the magnetic moment of
the neutrino.

The right-handed neutrino production
rate is proportional to the neutrino helicity flip probability,
which may be calculated by considering neutrino spin rotation in a
medium with an external magnetic field.
The complicated time
evolution of the neutrino spin can be described in
terms of a relativistic kinetic equation (RKE), which has been derived
in \cite{EnqvistSemikoz} and was extended to account
for the elastic collisions in
\cite{Semikoz1}.
The helicity rotation of a light Dirac neutrino traversing a magnetic field
is determined by
forward scattering off the field. Simultaneously, the neutrino
interacts with all the particles in the plasma via reactions that for light
neutrinos can be taken to conserve helicity.
 An essential feature in deriving the RKE is
averaging over the (random) magnetic field, the procedure of
which we improve in the present paper in order to discuss also
small scale fields. In what follows, we shall limit ourselves to the Standard
Model but assume that the neutrinos have Dirac masses.
%
%

If we neglect the magnetic moment  of the Standard Model Dirac neutrino,
in the ultrarelativistic limit the
dispersion relation, in hot plasma and in the presence of
a magnetic field, reads
$E(q)\approx q + V$. Here $V$
is the neutrino interaction potential which
is determined by the neutrino forward scattering amplitude and
which consists of two parts \cite{SemikozValle}:
\beq
V = V^{(vec)} + V^{(axial)}.
\label{potential}
\eeq
$V^{(axial)}$
includes contributions which are due to the fact
that in the presence of a magnetic field the charged background
is actually magnetized. This turns out to be an important effect.
The interaction potential $V^{(vec)}$ is generated by the mean vector
current $\sim \VEV{ \bar \psi_a\gamma_{\mu}\psi_a}_0$,
and at the temperature $m_e\ll T\ll\qcd$ it reads \cite{Notzold}
\beq
V^{(vec)}\approx 3.4\times 10^{-20}\Bigl (\frac{T}{\MeV}\Bigr )^5~\MeV~.
\label{vector}
\eeq
$V^{(axial)}$ is
generated by the mean axial current of the magnetized leptons
$\sim \VEV{
\bar{\psi}_l\gamma_{\mu}\gamma_5\psi_l}_0$ and is given by \cite{SemikozValle}
\beq
V^{(axial)} = \mu_{eff}\frac{{\bf q\cdot B}}{q} +
\frac{\mu_{eff}^2}{2q}\Bigl (
B^2 - \frac{({\bf q\cdot B})^2}{q^2}\Bigr )~,
\label{axial}
\eeq
where the quantity $\mu_{eff}$ is defined by\footnote{Note that this
effective magnetic moment has no relation with the anomalous neutrino
magnetic moment. }

\beq
\mu_{eff} = \frac{eG_F(-2c_A)T\ln 2}{\sqrt{2}\pi^2}\approx 6
\times 10^{-13}(-2c_A)\mu_B\Bigl (\frac{T}{\MeV}\Bigr )~,
\label{mueff}
\eeq
and  $c_A = \mp 0.5$ is the axial constant in the weak lepton current
(upper sign
for $\nu_{\rm e}$, lower for $\nu_{\mu,\tau}$).
Here $\mu_B = e/2m_e$ is the
Bohr magneton. In what follows
we may safely neglect the last term
in Eq. \eq{axial}.

The size of the random magnetic field domain $L_0$
influences crucially the neutrino spin behaviour.
For large-scale random magnetic field with $L_0\gg V^{-1}$
the dominant mode of evolution is
spin oscillation. For small scale magnetic fields (which is
perhaps a more realistic alternative in the early universe)
there also appears aperiodic spin motion which effectively converts
$\nu_L$ to $\nu_R$. We discuss $L_0$, dissipation, conductivity
and general constraints
on primordial magnetic fields in section 2. In section 3 we derive
a cosmological limit on the neutrino magnetic moment
in the case of a large scale magnetic field, taking into account
elastic collisions and improving on
previous treatments. Section 4 introduces an averaging procedure
which is suitable for small scale magnetic fields, and we provide
a limit on the neutrino magnetic moment
also in this case. Section 5 contains our comments and a discussion
of the meaning of the results.

\section{Direct constraints on magnetic fields}
 The early universe is an
excellent conductor, and a primordial magnetic field, if such existed at
any time, is imprinted on the comoving plasma which will retain
the field.
The magnetic flux is conserved so that the magnetic
field scales with the expansion of the universe as $B\sim R^{-2}$.
The strength of a random magnetic field  depends also on
how it scales over physical distances. If the field were to
perform random walk in 3-d volume, the scaling would be $B\sim N^{-3/2}$,
where $N$ is the number of steps. An argument based on the
statistical independence of conserved flux elements gives rise
to a $B\sim N^{-1}$ scaling. Whether the magnetic flux actually
is completely uncorrelated in the neighbouring unit cells is an unsolved
issue.
In \cite{eo} it was argued
that if the magnetic field domains are uncorrelated, a proper statistical
averaging produces a field with $\la B\ra =0$ and
a root-mean-square field $B_{rms}\equiv\sqrt{
B^2}\sim N^{-1/2}$. In the present paper we take a more phenomenological
view and assume merely that there exists a random magnetic field with
the scaling
\beq
B(L,t)=B_0\left ({R_0\over R(t)}\right )^2
\left({L_0\over L}\right)^p,                  \label{1}
\eeq
where $L$ and $L_0$ are two comoving physical scales, and $p$ is essentially an
unknown parameter; we shall focus mainly on the choices $p=1/2,\ 1,\ 3/2$.
We shall view $B(L,t)$ in Eq. \rf{1} as a root-mean-squared field with
$\VEV {B_j}=0$. We shall also assume isotropy so that $\VEV{B_iB_j}=0$
for $i\ne j$.
If we further assume that the magnetic domains are uncorrelated
we may write
\beq
\langle B_i({\bf x})B_j({\bf y})\rangle =
(2\lambda)^{-1}\delta_{ij}\delta^{(3)}({\bf x} -{\bf y})~,
\label{correlator}
\eeq
where the length $\lambda$ is determined by the domain size $L_0$ and
the rms field \rf{1} at the horizon scale $L = l_H(T)$:

\begin{eqnarray}
{\lambda^{-1}}&=& \frac{3}{\pi (3-2p)}B_{rms}^2(L=l_H)L_0^3~,~~p\neq 3/2~,\nn
{\lambda^{-1}}&=& \frac{3}{\pi}\Bigl (\ln \frac{l_H}{L_0}\Bigr )B_{rms}^2(
L=l_H)L_0^3~,~~p = 3/2~.
\label{lambda}
\end{eqnarray}

We may prove Eq. \rf{lambda}
by recalling the general fluctuation theory formula\footnote{For real
fields we use the + sign in the
$\delta$-function argument instead of the usual - sign for
the conjugated magnetic field components.}
\cite{Akhiezer1}
\beq
\VEV{ B_i({\bf k})B_j({\bf k}')} = (2\pi)^3\VEV{{\bf B}^2}_
{{\bf k}}\delta^{(3)}({\bf k} + {\bf k'})(\delta_{ij} - \hat {k}_i\hat {k}_j)~.
\label{fluct}
\eeq
If we substitute the Fourier transform $B_i({\bf x}) = \int d^3k\exp
(i{\bf kx})B_i({\bf k})/(2\pi)^3$ into the Anzatz \rf{correlator} and
use the rule Eq. \rf{fluct}, we obtain a constant spectrum
\beq
\VEV{{\bf B}^2}_{{\bf k}} =\frac{3}{2}\frac{1}{2\lambda} = {\rm const}~,
\label{spectr}
\eeq
where the factor $3/2$ arises because of a different
delta-function implementation in Eq. \rf{fluct}
as compared with Eq. \rf{correlator} .

Now we can express the parameter
$\lambda$ in Eq.  \rf{correlator} via the field in Eq. \rf{1}.
Let us first note that the {\em mean} magnetic field energy within a horizon
volume,
\beq
\rho_B = \frac{1}{V_H}\int d^3r\frac{B_{rms}^2(r)}{2} =
\frac {1}{2}\int \frac{d^3k}{(2\pi)^3}\VEV{
{\bf B}^2}_{{\bf k}}
\label{meandensity}
\eeq
depends\footnote{The second equality here has the sense of the mean field
energy density
$
\frac 12{\VEV{{\bf B}^2}_{{\bf x}=0}} =
\frac{1}{2}\int \frac{d^3k}{(2\pi)^3}\VEV{
{\bf B}^2}_{{\bf k}}
$
for fluctuating fields \cite{Shafranov}.
The first equality in Eq. \rf{meandensity} has this meaning as a
definition. Of course, for any
index $p$ in Eq. \rf{1} the magnetic field does not affect the expansion
of the universe because
$\rho_B\ll \rho_{\gamma}\sim T^4$. We use here the Gauss units in
the which for the uniform
field $\VEV{{\bf B}^2}_{{\bf k}} = B_0^2(2\pi)^3\delta^{(3)}({\bf k})$
the magnetic field energy density is given by
$\rho_B = B_0^2/2$.}
 on a cut-off wave number $k_{max}$:
\beq
\frac{2\pi}{V_H}B_0^2\Bigl (\frac{T}{T_{0}}\Bigr )^4\int_{L_0}
^{l_H}r^2dr\Bigl (\frac{L_0}{r}\Bigr )^{2p} =
\frac{k^3_{max}}{16\lambda\pi^2}
{}~.
\label{connection}
\eeq
Physically $k_{max} = 2\pi/L_0$ corresponds to a minimum
size of the inhomogeneity, and is given as in
\cite{EnqvistSemikoz} by
\beq
\frac{3}{5}k_{max}^2 =
\frac{\int k^2d^3k\VEV{{\bf B}^2}_{{\bf k}}/(2\pi)^3}{\VEV{{\bf B}
^2}_{{\bf x}=0}}~.
\label{kmax}
\eeq
Here the factor $3/5$ arises by using Eqs. \rf{spectr} and \rf{meandensity}.

Substituting Eq. \rf{kmax} into Eq. \rf{connection} we arrive at the
 relations Eq. \rf{lambda} between
$\lambda$, the domain size $L_0$ and the field $B_{rms}$ at the horizon scale
$l_H$:
\beq
{\lambda^{-1}} = \frac{3}{\pi (3 - 2p)}B_0^2\Bigl (\frac{T}{T_{0}}\Bigr )^4
\Bigl
(\frac{L_0}{l_H}\Bigr )^{2p}L_0^3 = \frac{3}{\pi(3 - 2p)}B_{rms}^2(L= l_H)
L_0^3~,~~p\neq 3/2~,
\eeq
and similarly for $p = 3/2$.

Let us now turn to the constraints on the domain size $L_0$
and on the magnetic field strength in the plasma of the early universe.
We begin by discussing the dissipation of primordial magnetic fields.

Conductivity in the early universe, although large, is not infinite,
and accordingly there will be some ohmic dissipation, starting from small
length scales, $L_0\leq L_{diss}$ . The dissipation scale is given by
\beq
L_{diss} = \sqrt{\frac{t}{4\pi \sigma_c}}~,
\label{dissipation}
\eeq
where $t$ is the age of the universe and $\sigma_c$ is conductivity, which
for isotropic relativistic electron gas reads
\beq
\sigma_c = \frac{\omega_p^2}{4\pi\sigma n_e} = \frac{T}{\pi\alpha L_c}~.
\label{conductivityrel}
\eeq
Here $\omega_p = \sqrt{4\pi \alpha n_e/T}$ is the plasma frequency and
$\sigma\approx \pi \alpha^2L_c/T^2$ is the collision cross section and $L_c\sim
3$ is the Coulomb logarithm. We then find that the dissipation length below
$T_{\rm QCD}$ is given by
\beq
L_{diss}^{rel} = 0.1g_*^{-1/4}\Bigl (\frac{\MeV}{T}\Bigr )^{3/2}~\cm~.
\label{dissipation3}
\eeq
The finite isotropic conductivity
\rf{conductivityrel} reflects the
properties of the plasma at large scales when we may neglect the
random magnetic field influence upon relativistic plasma.  We may thus
use the estimate \rf{dissipation3} to obtain a conservative lower bound
on the domain size $L_0$.

It is worth noting that within a domain volume with a uniform magnetic field
where plasma is highly anisotropic, the dissipation length is greatly
reduced with
respect to Eq. (17) because of substantial
increase of conductivity, i.e. a strong
magnetic field tends to sustain itself within a given domain.
Indeed, for relativistic electron-positron plasma in a strong uniform
magnetic field, which is much larger than the quantizing magnetic field
$B_c = m_e^2/e = 4.41\times 10^{13}~{\rm G}$, conductivity is given
by
\beq \sigma_c = \frac{\omega_p^2}{4\pi\sigma n_e} = \frac
{T(B/B_c)}{4\pi \alpha C_E^2}~.  \label{conductivity}
\eeq
Here $\sigma
\approx 4C_E^2\pi\alpha^2(B_c/B)/T^2$ is the $e^+e^-$ collision cross
section \cite{Langer} for the electrons and the positrons that
occupy the lowest
Landau level, and $C_E\approx 0.577$ is the Euler constant. The
dissipation length within a domain with a strong uniform field is
given by \beq L^{anis}_{diss} = \frac{3.5\times 10^{-3}~{\rm
cm}}{g_*^{1/4}} \Bigl (\frac{\MeV}{T}\Bigr )^{3/2}\times \Bigl
(\frac{B_c}{B}\Bigr )^{1/2}~.  \label{dissipation1} \eeq Let us note
that within one domain with a uniform magnetic field the
ratio $B/B_c$ can be interchanged with the temperature ratio,
$B/B_c\sim (T/{\rm MeV})^2$. The dependence of the magnetic field on the
temperature reflects the magnetic flux conservation. Here
we neglected the small contribution to the plasma conductivity produced by the
collisionless
one-photon processes in a strong external magnetic field
($\gamma\leftrightarrow e^+e^-,~~~e^{\pm}\leftrightarrow e^{\pm}\gamma$). This
conductivity is calculated from the imaginary part of the polarization
tensor in magnetic field \cite{Perez} and can be estimated as
\beq
\sigma^{one-photon}_c \simeq \frac{\alpha(B/B_c)m_e^2}{4\pi T}\ll
\sigma_c~,
\eeq
where $\sigma_c$ is given by Eq. \rf{conductivity}.

Comparing the relativistic dissipation length
\rf{dissipation3} with the analogous result for anisotropic plasma
\rf{dissipation1} we find that one can neglect the magnetic field dissipation
at small scales since domains with
sizes $L_0(T)\geq L_{diss}^{rel}(T)\gg L_{diss}^{anis}$ survive.

In the non-relativistic case, which is relevant for the recombination time
$t\simeq
10^{12}~{\rm sec}$ ($T_{rec} \sim 0.4~{\rm eV}$) and for the BBN time
$t\sim 1~ {\rm min}$
($T_{BBN}\sim 0.1~{\rm MeV}$), one obtains a different
expression for the dissipation length,
\beq
L^{non-rel}_{diss} = 0.1g_*^{-1/4}\Bigl (\frac{\MeV}{T}\Bigr )^{7/4}~\cm~,
\label{dissipation2}
\eeq
where we have used the conductivity of non-relativistic isotropic hydrogen
plasma \cite{Akhiezer}
\beq
\sigma_c=\frac{ 4\sqrt{2}T^{3/2}}{\pi^{3/2}\alpha L_cm_e^{1/2}}~.
\label{conductivity2}
\eeq
The difference between this and the relativistic expression
\rf{conductivityrel}
follows from: (i) the Debye screening for the
forward $ee$ or $ep$ scattering in isotropic non-relativistic plasma
with the Coulomb logarithm $L_c\sim 10$, and (ii) the fact that in the
non-relativistic case the velocity of the electrons is the thermal velocity
 $v_{T_e} = \sqrt {T/m_e}\neq 1$.

The dissipation length \rf{dissipation2} at the recombination time
is about $\sim 10^{10}~{\rm cm}$ which
translates to the BBN time $t\sim 1~ {\rm min}~(T_{BBN}\sim 0.1~{\rm MeV})$ as
$L^{min}_0(T_{BBN}) = 10^{10}~\cm\times (T_{rec}/T_{BBN})\simeq
4\times 10^4~\cm$.
The domains which at that time are larger than this scale,
$L_0\gsim L_0^{min}(T_{BBN})$, survive after the recombination time, and
such a relic field could be a seed field
for the dynamo enhancement of the galactic
magnetic field.
This requirement can be rewritten as (see also \cite{Cheng})
\beq
L_0\geq L_0^{min} = 10^3~{\rm cm}\times \Bigl (\frac{\MeV}{T}\Bigr )~.
\label{domainsize}
\eeq

Let us note that the {\it
local} dissipation length calculated at the same BBN time from
Eq. \rf{dissipation2}
is significantly smaller, $L_{diss}^{non-rel}(T_{BBN})\sim 5~{\rm cm}$, i.e.
domains with
sizes within the region $5~{\rm cm}\lsim L_0(T_{BBN})\lsim L_0^{min}(T_{BBN})
\sim 10^4~{\rm cm}$
continue to expand after the BBN time but they dissipate before the
recombination
time and do not contribute to the relic seeding of galactic magnetic field.

For instance,
the relativistic dissipation length \rf{dissipation3}
at the temperature corresponding to the electron neutrino decoupling
with matter, $T\sim T_d\sim 2~\MeV$,
is even smaller,
\beq
L^{rel}_{diss}(T\sim 2~\MeV)\sim 10^{-2}~{\rm cm}~,
\label{reldis}
\eeq
or $\sim 0.3~{\rm cm}$ after its expansion till the BBN time.
Therefore, all the domains that have
survived in the hot plasma but have at $T_{BBN}$ a size within
the range $0.3~{\rm cm}\lsim L_0(T_{BBN})\lsim 5~{\rm cm}$,
in fact dissipate before $T_{BBN}$.
On the other hand, all the domains that survive after the nucleosynthesis but
have a size less than $10^4~\cm$ at $T_{BBN}$ dissipate before the
recombination
time.

Accounting for the expansion of the domains, we find that if
\beq
L_0(T)\gsim L_0^{min'}=10^{-2}~{\rm cm}\times \Bigl (\frac{\MeV}{T}\Bigr )~,
\label{domainsize2}
\eeq
such domains are guaranteed to survive at any relativistic temperature;
however, they dissipate before the BBN time if their scale $L_0(T_{BBN})$
was less than $L_{diss}^{non-rel}(T_{BBN})= 5~\cm$.

Primordial nucleosynthesis considerations can further be used to constrain the
strength of the magnetic field. If large enough, a magnetic field would affect
the expansion rate of the universe as well as the rates of the
various reactions that are involved in building up the abundances
of light elements.
Including these effects in a nucleosynthesis code
Cheng et al. find in a recent study \cite{Cheng}  an upper bound
\beq
\ns\equiv B(10^4\;\cm ,1\;\min )< 10^{11}\G .      \label{2}
\eeq
This is a very useful constraint, as will become evident shortly.

The energy density in the magnetic field within the causal horizon is given by
\beq
\rho_B ={2\pi B_0^2\over V_H}\left (\frac{R}{R_0}\right)^4
\int_{L_0}^{l_H}drr^2\left( {L_0\over r}
\right)^{2p},      \label{3}
\eeq
where $V_H$ is the causal  horizon volume.
(At $\ew\simeq 100$ GeV the causal horizon size is $l_H\simeq 1.4$ cm).  A
reasonable requirement is that the energy of the magnetic field fluctuations
should not exceed the free electromagnetic energy, so that $\rho_B\lsim
T^4$. This is not in disagreement with the nucleosynthesis constraint
Eq. \rf{2} at any scale below $\ew$. If we assume, as is natural, that
the random field fluctuations are bounded by $B(L_0,t)\lsim {\cal O}
(T^2)$, then $\rho_B\lsim T^4$ for $l_0<l_H$ for all $p$ under consideration.

If we assume that the observed galactic magnetic fields are due to a
large scale dynamo effect \cite{dynamo}, seeded by the primordial field,
then we may
set a lower bound on the strength of the cosmological seed field $\seed$
(this is further enhanced by a factor of $10^4$ by the collapse
of the protogalaxy) \cite{Shukurov}.
Numerical simulations of the
field growth appear to imply that the seed field must be sufficiently
large for the dynamo to work \cite{brand}, and
one should require that
at $t\simeq 10^{10}\yr$ and at the intergalactic scale $\gal=100$
kpc, $\seed\equiv
B(\gal, 10^{10}\yr )\gsim 10^{-18}\G$. This bound can be made even tighter
provided we assume that the observed field reversal between the Orion
and Sagittarius arms \cite{orion} is related to the dynamo.
To produce a field reversal, the seed field should have been relatively strong
\cite{anvar}, $\seed\gsim 10^{-11}\G$. Such reversal has only been observed
in the Milky Way and may thus not be a generic feature.

{}From Eq. \rf{2} it follows that
\beq
\seed=\ns \left({1 {\rm min}\over t_*}\right)\left({t_*\over t_{now}}
\right)^{4/3}\left({10^4{\rm cm}\over \gal}\right)^p<
1.8\times 2.4^p\times 10^{-7-11p}\G,            \label{5}
\eeq
where we have assumed that the change to matter
dominated universe  takes place at $t_*\simeq 8750$ yr, corresponding
to $\Omega h^2\simeq 0.4$. Therefore, the case
$p=3/2$, which translates into the bound
$\seed <2\times 10^{-23}\G$, cannot provide a seed field large
enough for the galactic dynamo, no matter what mechanism generated the
field in the first place. If $p=1$ we obtain $\seed < 4\times 10^{-18}$ G but
given the large theoretical uncertainties, this case could still be
compatible with the galactic dynamo.
Finally, if the observed field reversals in the Milky Way
are really related to the dynamo, then only the $p=1/2$ case remains marginally
compatible.

It has been argued \cite{Hector}
that in magnetic fields of the order of $B\approx 10^{18}$
G the neutron becomes stable against $\beta$-decay and that for somewhat
larger fields proton becomes unstable to a decay into a neutron.
If $B\lsim T^2$ at all scales, this effect would not be important
for nucleosynthesis. It is also irrelevant for the problem of
excitation of the wrong-helicity neutrino states because
neutrino spin flip is determined by the field
strength at $\qcd$ and at scales of the order of the weak collision
length $L_W$. Indeed, we may write the
the nucleosynthesis bound \rf{2} as
\beq
B(\lw ,\qcd)\lsim 8\times 10^{17}\times 13^{-p}\G~\lsim ~2\times 10^{17}\G
\label{6}
\eeq
Here we have used the estimate $L_W\approx 1.6\times 10^{-4}$ m.
Considering the inherent uncertainties, we may then well adopt the
latter figure in Eq. \rf{6} as a conservative bound at $\qcd$ for all $p$.

Finally, note that the evolution of a magnetic field in an expanding Universe
is determined by
MHD which should explicitly
yield the domain size $L_0$ as well as the
topological index $p$ in Eq. \rf{1}, both of which are crucial for the neutrino
propagation
in a medium with a random field. We shall use the phenomenological model
\rf{1} for case of uncorrelated 3-d fields in
Eq. \rf{correlator} with an arbitrary index $p$.

%
\def\bef{\begin{figure}}
\def\eef{\end{figure}}
\def\bet{\begin{table}}
\def\eet{\end{table}}
\def\bea{\begin{eqnarray}}
\def\ba{\begin{array}}
\def\ea{\end{array}}
\def\bi{\begin{itemize}}
\def\ei{\end{itemize}}
\def\ben{\begin{enumerate}}
\def\een{\end{enumerate}}
\def\ot{\otimes}

\section{Neutrino propagation in medium with large-scale random magnetic field
}
Neutrino spin-flip in a magnetic field
can affect the Big Bang nucleosynthesis
 of light elements because of the appearance of an additional
gravitating relativistic component in the plasma.
The nucleosynthesis limit on the extra degrees of
freedom at the time of nucleosynthesis is \cite{OS},
in units of relativistic two-component
neutrinos, $\Delta N_{\nu}\lsim 0.1$ (see also \cite{Schramm}).
Indeed,
the wrong--helicity neutrinos will be abundant at the time when
the neutron--to--proton ratio freezes at $T\simeq 0.7$ MeV,
violating the nucleosynthesis bound, unless
they decouple before the QCD phase transition. Then their relative number
densities will be diluted to acceptable levels by the heating of
the particles still in equilibrium.

In the early universe
the production rate of the wrong--helicity neutrinos is given by
$\Gamma_{L\to R}=\la P_{\nu_L\to\nu_R}\ra{\tot}$, where
$\tot$  is the total weak collision rate, and in the absence of
collisions the averaged conversion probability reads
\beq
\la P_{\nu_L\to\nu_R}\ra=\frac 12{\hh^2\over \omega^2},  \label{110}
\eeq
where $\hh = 2\mu_\nu B_\perp$ is the field perpendicular to the
neutrino propagation, $\mu_\nu$ is the neutrino magnetic moment and
\beq
\omega^2=2\hh^2+V^2+6L_0^{-2}/5
\label{omega2}
\eeq
is the spin rotation frequency. Note that it depends both on the value
and scale of the random field. In a uniform constant field one would
find $\omega^2=\hh^2+V^2$ in the expression \rf{110}. In a realistic case
$V\gg\hh$.

To find out $\Gamma_{L\to R}$,
let us consider the electron neutrino for
definiteness. Then at $T=\qcd$ one should include
the following processes: $
\nu_{\rm e}\bar{\nu}_{\rm e}\to l^+l^-,\ \nu_\alpha\bar{\nu}_\alpha,\
 q\bar{q}\; ;\ \nu_{\rm e}l^{\pm}\to\nu_{\rm e}l^{\pm};\
\nu_{\rm e}{\nu}_{\alpha}
\to\nu_{\rm e}\nu_\alpha ;\ \nu_{\rm e}\bar{\nu}_{\alpha}
\to\nu_{\rm e}\bar{\nu}_\alpha;\ \nu_{\rm e}q\to\nu_{\rm e}q\; ;\
\nu_{\rm e}\bar{q}\to\nu_{\rm e}\bar{q}\; ;\ \nu_{\rm e}e^+\to u\bar{d}\; ;\
\nu_{\rm e}\mu^{-}\to\nu_{\mu}\mu^{+};\ \nu_{\rm e}\bar{\nu}_{\mu}\to
\mu^{-}e^{+}.$ Here the notation is: $\alpha={\rm e},\ \mu ,\ \tau;\ l={\rm
e},\
\mu\; ;\ q=u,\ d$.
Adding up all these processes we find the
thermally averaged electron neutrino collision rates to be
\beq
\tot (\qcd )\approx 30G_F^2\qcd^5;~~ \Gamma_W^{\rm el}(\qcd )\approx
2G_F^2\qcd^2.
\label{14}
\eeq
Here we have neglected the decays and inverse decays, whose contributions
are small. The main contribution to Eq. \rf{14} comes from the charged current
processes involving quarks. For $\nu_\mu$ and $\nu_\tau$ the rate
is slightly different.

Requiring $\Gamma_{L\to R}\lsim H$
at $T=\qcd$ where $H=\sqrt{
8\pi^3g_*(T)/90}T^2/M_{Pl}$ is the Hubble parameter, one obtains
a constraint
on the product of the Dirac neutrino magnetic moment and the mean squared
random field $B= \langle B^2\rangle^{1/2}$ \cite{EOS}.
Here we adopt the value $g_*(\qcd )\simeq 63$, which includes
also the effects due to the non--relativistic species \cite{uibo}.
The limit thus obtained does not depend on any model for the random field, and
one finds that
\beq
\mu_{\nu}B(\qcd,L_W)\lsim 4\times 10^2\mu_B\Bigl (\frac{\qcd}{200~\MeV}\Bigr
)^{7/2}~\G~.
\label{limit1}
\eeq
The result \eq{limit1} is true in {\it collisionless} regime if the random
domain size
$L_0$ is larger than the
neutrino spin oscillation length $l_{osc} \simeq V^{-1} $ with
\beq
l_{osc}=10^{-2}l_H(T/\MeV)^{-3}~,
\label{Lmin}
\eeq
where $l_H\sim M_{Pl}/T^2$ is the horizon length.
For small domains, $L_0\ll l_{osc}$, neutrino spin cannot
follow the random direction of the magnetic field and the neutrino spin
rotation effectively ceases \cite{EnqvistSemikoz}. The limit \eq{limit1}
is slightly different from the one presented in \cite{Shukurov} because of
the inclusion of the quarks in the total collision rate.

If we take into account {\it elastic} neutrino collisions with charged
particles in the background
plasma, the analogous consideration starts from
the RKE for the $z$-component of the neutrino spin $S_z(t)= 2P_{\nu_L
\leftrightarrow \nu_R}(t) - 1$. This is an
integro-differential
equation and given by \cite{Semikoz1}
\bea
\frac{dS_z(t)}{dt} =& -2\int_0^t\exp [-\int_{t_1}^{t}\nu_{\perp}(t_2)
dt_2]\times \Bigl [\tilde{H}_-(t)\tilde{H}_+(t_1)\exp
(i\int_{t_1}^{t}V(t_2)dt_2) \nn
&+ \tilde{H}_+(t)\tilde{H}_-(t_1)\exp (-i\int_{t_1}^{t}V(t_2)dt_2)\Bigr ]S_z
(t_1)dt_1~, \label{spineq}
\eea
where $\tilde{H}_{\pm}(t) =
\mu_{\nu}(B_x(t) \pm iB_y(t))$~
depends on the transversal field
components only. The collision frequency $\nu_{\perp}(t)$
is approximately given by the
weak elastic collision rate $\Gamma_W^{el}$.
In the absence of {\it inelastic} collisions
the relaxation of the transversal spin components
is completely determined by Eq. \eq{spineq}.
If
$V$ and $\nu_{\perp}$ are
slowly varying, one may transform Eq. \eq{spineq}
to a differential equation
of third order, after which one may perform the averaging over
the random magnetic field by assuming isotropy and using the
spectral density representation
for the magnetic field correlators. One finds \cite{Semikoz2}
that
the probability for helicity change is given by
\bea
P_{\nu_L\to\nu_R}(t)&=&{2\hh^2\over \om^2}[1-\exp (-\nu_\perp t)(\cos \om t
+{3\nu_\perp \over\om}\sin \om t)] \nn
&+&\frac 12 [1-\exp\left(-{8\hh^2\nu_\perp  t\over \om^2}\right)].  \label{10}
\eea

Note that in the presence of elastic collisions the spin-flip probability
increases and tends towards the asymptotic value 1/2 as $t\to\infty$.
In the present section we assumed that all the collisions take place
within a homogenous magnetic domain. Hence,
when evaluating the production rate of right-helicity neutrinos, one should
calculate the probability when $t\simeq L_0$. During this time the
neutrino has been subject to a large number of independent collisions,
each of which have served to adjust the spin-content of this state.
The scale of the magnetic field felt by the neutrino at each collision
is given by the free path length $L_W$. Thus we find
\beq
\Gamma_{L\to R}={4\hh^2\nu_\perp L_0\tot\over V^2}.
\label{GLR1}
\eeq
Assuming $\nu_\perp\approx \el$ it is then straightforward to deduce
 a constraint on the product $\mu_{\nu}B$:
\beq
\mu_{\nu}B(\qcd,\lw )\lsim 3.5\times 10^2\left({\lw\over L_0}\right)^{1/2}~.
\label{limit2}
\eeq
Note that by taking into account elastic collisions
one obtains a limit
which is more stringent than in the collisionless case.
This is due to the fact that spin rotation turns the longitudinal
part of the spin into transversal, and
at each collision the transversal
part is, in effect, wiped out. This results in a shrinkage of the
spin vector and as $t\to\infty$, $P_{\nu_L\to\nu_R}(t)\to 1/2$.
Qualitatively one can also see this in the following manner.
Let us write the RKE in the familiar form
\beq
\frac{d{\bf S}}{dt}={\bf V}\times{\bf S}-\nu_\perp{\bf S}_\perp,
\eeq
where now ${\bf V}=V\hat{n}_z+2\mu_\nu(B_x\hat{n}_x+B_y\hat{n}_y)$
and the transversal spin is given by $S_\perp\sim S(V_\perp/V_z)$
with $V_\perp = 2\mu_\nu B_\perp$.
It then follows that
\beq
{dS^2\over dt^2}=-2\nu_\perp S^2_\perp,
\eeq
yielding a shrinkage rate $S^{-2}dS^2/dt=-\nu_\perp\times 8\la
\tilde{H}^2_\perp\ra /V^2$, in agreement with Eq. \rf{10}.

The usefulness of the limit Eq. \eq{limit2} in restricting
the neutrino magnetic moment depends of course on the magnitude
of the primordial magnetic field. Adopting the maximum value
allowed by nucleosynthesis, given in Eq. \rf{6}, and taking
$L_0\approx l_H$ we see that the tightest possible constraint which
in principle can be obtained in this manner is $\mu_\nu\lsim 3\times 10^{-19}
\mu_B$.
 Cosmological \cite{cosmo}
and astrophysical \cite{raffelt} constraints on $\mu_\nu$,
based on the direct production of wrong-helicity neutrinos
in photon mediated collisions are typically less severe by several
orders of magnitude. Thus the presence of a primordial magnetic field
is a potential bonus for neutrino physics.

The  derivation of Eq. \eq{limit2} assumes that $L_0
\gg \Gamma_W^{-1}$. This means that a scattered neutrino meets
always a transversal part of a randomly orientated magnetic field.
It seems not very likely, however, that the coherence length of the
magnetic field at $T=\qcd$ could be as large as the horizon size,
especially if the origin of the field is at earlier times when
the size of the horizon was much smaller. Thus,
in the next section we consider the corresponding limit in the
case of a small scale magnetic field.
\section{ Neutrino propagation in medium with small-scale random
magnetic field}

An important technical point in the derivation of
Eqs. \eq{limit1} and \eq{limit2}
is the procedure by which one
averages the spin equation of motion over the random magnetic field
distribution.
For the limits \eq{limit1} and \eq{limit2} the exact differential
equation derived from Eq. \eq{spineq} was averaged after
it was first transformed
to a more suitable form.
Such a procedure is always valid for a regular
magnetic field, but in the case of random fields we should be more careful.
In fact, as we shall now show,
in the case of small domains with $L_0\ll l_{osc}$ there appears aperiodic
neutrino spin motion.

Let us rewrite Eq. \eq{spineq} as
\bea
\frac{dS_z(t)}{dt} =& -4\int_0^t\exp [-\int_{t_1}^{t}\nu_{\perp}(t_2)
dt_2]\times \Bigl [{\rm Re~} \Bigl (\tilde{H}_-(t)\tilde{H}_+(t_1)\Bigr )\cos
(\int_{t_1}^{t}V(t_2)dt_2) \nn \\
&+ {\rm Im~} \Bigl (\tilde{H}_+(t)\tilde{H}_-(t_1)\Bigr )\sin
(\int_{t_1}^{t}V(t_2)dt_2)\Bigr ]S_z
(t_1)dt_1~.
\label{spineq2}
\eea
Assuming that the collision frequency $\nu_{\perp}$ depends only
weakly on the magnetic field,
and taking into account that in the leading approximation the potential
$V(t)$ as given in Eq.
\eq{potential}  depends  on $B_z(t)$ only while $\tilde{H}_{\pm}
(t)$ are proportional to the transversal components, we can average these
factors in integrand {\it independently} because of the isotropy of the system.
For neutrinos crossing many small-scale domains with
$t\gg L_0$, the size $L_0$ corresponds to a narrow resonance
for uncorrelated random fields, as we now show.

To this end, consider the kernel
\beq
K(t - t_1) = \VEV{\tilde{H}_+(t)\tilde{H}_-(t_1)}/\VEV{
\tilde{H}_{\perp}^2}
\eeq
 with
\beq
\VEV{\tilde{H}_{\perp}
^2}= (2/3)\mu_{\nu}^2B^2.
\eeq
If the fields are uncorrelated\footnote{Here we consider one-dimensional
correlators, but the same result could be obtained by using the
full 3-d correlator given in Eq. \eq{correlator}.
There the factor $\lambda$  is  related to the horizon scale because
$\la {\bf B}^2\ra$ involves integration  over all space.
}
 with $\VEV{B(t)B(t_1)}=B^2L_0\delta (t-t_1)$ one
finds
\beq
\frac{K(t)}{L_0}\sim \lim_{L_0\rightarrow 0}\frac{L_0}{t^2 + L_0^2} =
\frac {\pi}{2}\delta (t)~.
\eeq
For such uncorrelated fields the averaging over
of the transversal components then results in $\delta (t - t_1)$ under the
integral in Eq. \eq{spineq} which wipes out the exponent in the integrand
and leads (for the initial condition $S_z(0)= -1 $)
to a new damping solution
\bea
S_z(t) &=& - \exp (- \Gamma t)~,\nn
P_{\nu_L\rightarrow \nu_R}(t) &=& \frac{(1 - \exp(-\Gamma t))}{2}\approx
\Gamma t/2~,    \label{conversion}
\eea
where the damping parameter $\Gamma$ is given by
\beq
\Gamma = \frac {8}{3}\mu_{\nu}^2B^2L_0~.
\label{damping}
\eeq
One should bear in mind that the averaged field $B$ depends on the horizon
scale
$L = l_H$ and the domain size $L_0$ for uncorrelated fields (see
Eq. \rf{correlator}).

As in the previous section, the spin-flip probability grows with time.
Now we assumed that $\lw\gg L_0$ so that we should calculate the probability
at largest possible time.
Setting $t = H^{-1}$ we find
\beq
\Gamma_{L\to R}=\frac 43\mu_\nu^2B^2L_0H^{-1}\tot.
\label{GLR2}
\eeq
Note that $\Gamma_{L\to R}\sim T^3$, so that again we should require
$\Gamma_{L\to R}<H$ at $T=\qcd$. This results in the bound
\beq
\mu_\nu B(\qcd,l_H)\lsim 6.7\times 10^{-3}\mu_B\G~\left({\lw\over L_0}
\right)^{1/2}.
\label{limit3}
\eeq
Note that this bound does not agree with Eq. \eq{limit2} in the limit
$L_0\to\lw$. This is due to the qualitatively different averaging procedures
at large and small $L_0$. When $L_0\approx\lw$, neither method is reliable.

Substituting Eq. \rf{1} to Eq. \rf{limit3} we can rewrite the BBN constraint
on the Dirac neutrino magnetic moment as
\beq
\mu_{\nu}\lsim \frac{1.7\times 10^{-21}\mu_B}{(L_0^{min})^{p +
1/2}\Gamma_W^{1/2}
H^p}~,
\label{moment}
\eeq
where the minimum domain scale $L_0^{min}(T_{QCD})$, the left neutrino total
collision rate $\Gamma_W(T_{\rm QCD})$ as given by Eq. \rf{14},
and the Hubble parameter $H(T_{\rm QCD})$ are the
functions
of the temperature $T_{\rm QCD} = 200~\MeV\times T_{200}$.

As we have discussed
in section 2 there are two scenarios for the choice of the minimum scale
$L_0^{min}(T)$. Let us first assume that the relic field
is the seed for the galactic magnetic field, so that the
domains with a size $L_0\geq L_0^{min}(T) = 10^3~\cm(\MeV/T)$
survive after the recombination time. In the second scenario\footnote{The
first
scenario is not necessary since there are other possibilities for the seed
field creation in the MHD-dynamo theory of the galactic magnetic fields.} ,
with the use of the
relativistic plasma dissipation length \rf{dissipation3}, the domains with
sizes $L_0\geq
L_0^{min'}= 10^{-2}~\cm(\MeV/T)$ survive at any relativistic temperature
$T\gsim
2~\MeV$, but dissipate even before the BBN temperature $T\sim 0.1~\MeV$.
This last fact
does not matter for the neutrino spin-flip that populates wrong helicity states
mainly around $T\sim \qcd$. Thus, we find from Eq. \rf{moment} two upper limits
for the Dirac neutrino magnetic moment:
\beq
\mu_{\nu}\lsim \frac{10^{-22 + 6p}\mu_B}{(5.4)^pT_{200}^{p + 2}}~,
\label{limit4}
\eeq
for the first case with the relic seeding of galactic magnetic fields, and
\beq
\mu_{\nu}\lsim \frac{3.2\times 10^{-20 + 11p}\mu_B}{(5.4)^pT_{200}^{p + 2}}~,
\label{limit5}
\eeq
for the second scenario with dissipation of the random fields before the BBN
time.

As an example, for the index $p = 1/2$ we obtain from Eqs.
\rf{limit4} and
\rf{limit5} very restrictive constraints on the Dirac neutrino magnetic
moments,
$\mu_{\nu}\lsim 4.3\times 10^{-20}\mu_B/T_{200}^{5/2}$, or $\mu_{\nu}\lsim
4.4\times 10^{-15}\mu_B/T_{200}^{5/2}$, respectively.
These numbers are deduced by requiring that there should not occur full
equilibration of one right-handed neutrino species.
If we require that $\Delta N_{\nu}\lsim 0.1$, we should
multiply the upper limits above by a factor
$(\Delta N_{\nu})^{-1/2}\sim 3$.  However, we wish to emphasize
that the very stringent  constraints
on the Dirac magnetic moment above are very sensitive to the model of
the primordial magnetic field.
We should also point out that the both scenarios are based on the
common assumption that the magnetic coherence length
is much larger than the interparticle distance, $L_0\gg T^{-1}$,
an assumption which is natural considering the
macroscopic nature of the
"glueing" of the magnetic field force lines on hot plasma.

The remaining issue is the
validity of the RKE Eq. \eq{spineq} when also inelastic collisions
are taken into account. This work is now in progress.

\section{Discussion and conclusions}

We have found out that the general constraints on the primordial magnetic
fields, as implied by nucleosynthesis, do not exclude
the possibility of very tight limits
on the Dirac neutrino magnetic moments
in the presence of magnetic fields. These can be derived from the
requirement that the right-handed components should not be in
thermal equilibrium at time of
nucleosynthesis. We found, however, that nucleosynthesis and the
dynamo origin of the observed galactic magnetic fields are not compatible
with a magnetic field model consisting of uncorrelated cells with a scaling
index
$p=3/2$, and only marginally compatible if
$p=1$. If the field reversals observed in the Milky Way
are due to the dynamo, then
only the scaling law $p\lsim 1/2$ is appropriate (this naturally includes
the constant background $p=0$ as in the case of the Savvidy vacuum
\cite{poul}).

The particle physics aspect of the magnetic moment constraints is
straightforward: one only needs to know the
neutrino collision rates just above the QCD phase transition.
Neutrino spin evolution is then determined by a general RKE.
Here we included only the effects due to the elastic scattering.
Generally speaking, we should also account
for the dependence of the weak rates on
the strong magnetic field.
For the spin collision integrals this problem is now in progress,
together with the generalization of the RKE \eq{spineq} to the case
of inelastic collisions. It is not obvious how ineleastic collisions
affect the evolution of the neutrino spin.

The actual limits on the neutrino magnetic moments depend on two unknowns:
the strength of the field at $T\approx \qcd$, and the size of the homogenous
magnetic field domain $L_0$. The magnitude of the field depends on the
mechanism by which it was first produced, at the electroweak phase transition
or earlier, and on its scaling when averaged over several domains,
if $L_0\ll \Gamma_W^{-1}$. In a given model of the primordial magnetic
field both these can be estimated, but
here we adopted a phenomenological view
and simply assumed $B\sim (L_0/L)^p$. The magnitude of $L_0$ is a more
complicated issue as it involves magnetohydrodynamics in the hot plasma
of the early universe. It seems obvious, though, that $L_0\gg 1/T$ as
the build-up of the magnetized plasma requires a large number of particles,
but that $L_0$ is (much) less that the horizon size. One possibility would
be that the collision frequency
of charged particles (i.e conductivity) plays a decisive role in
forming homogenous regions. This remains an open problem. In the present
work we studied the cases with $L_0\ll \Gamma_W^{-1}$ and $L_0\gg
\Gamma_W^{-1}$, which lead to qualitatively different evolution equations.
We also gave some examples of the possible order of magnitude of the
upper limits on $\mu_\nu$ which could turn out to be as restrictive
as $10^{-20}\mu_B$. The limit also depends on whether
the primordial field provides the seed for galactic dynamo,
and thus in the absence of a more detailed knowledge
of the dynamics of the primordial magnetic fields no definite statement
about the actual constraint on neutrino magnetic moments
can be made. We may however conclude that potentially the
primordial magnetic field constraint on Dirac neutrinos could be
very important.

\vskip 1truecm

{\bf Acknowledgements}
A. Rez would like to thank The Education and Research Center "Cosmion"
for initiative grant. V. Semikoz acknowledges the High Energy Physics group
of the Research Institute for Theoretical Physics at the Helsinki University
for hospitality during the preparation of this work.
 We thank J. Maalampi  for fruitful discussions.

\end{document}